\documentclass[11pt,a4paper]{article}
\usepackage{amssymb,amscd}
\title{Algebraic curves, integer  sequences and a discrete Painlev\'{e} 
transcendent} 

\author{
A.N.W. Hone \\
\normalsize
\em  Institute of Mathematics, Statistics \&
Actuarial Science, \\
\normalsize
\em
University of Kent, \\
\normalsize
\em Canterbury CT2 7NF, UK.\thanks{
E-mail: anwh@kent.ac.uk \newline
2000 MSC class: 11B37, 33E05, 37J35.} \\ 
{\bf Poster at Symmetries \& Integrability of Difference Equations,} 
\\ 
{\bf SIDE 6, 
Helsinki, Finland, 19-24 June 2004.}  
}
\begin{document}
\renewcommand{\theequation}{\arabic{section}.\arabic{equation}}
\newcommand{\beq}{\begin{equation}}
\newcommand{\eeq}{\end{equation}}
\newcommand{\bea}{\begin{eqnarray}}
\newcommand{\eea}{\end{eqnarray}}
\newcommand\la{{\lambda}}
\newcommand\ka{{\kappa}}
\newcommand\al{{\alpha}}
\newcommand\be{{\beta}}
\newcommand\si{{\sigma}}
\newcommand\lax{{\bf L}}
\newcommand\mma{{\bf M}}
%\renewcommand{\thepage}{}
%\begin{titlepage}
\Large

\maketitle

\begin{abstract}
We consider some bilinear recurrences that have applications in 
number theory. The explicit solution of a general three-term bilinear
recurrence relation of fourth order
is given in terms of the Weierstrass sigma function for an 
associated elliptic curve. The recurrences can generate 
integer sequences, including the Somos 4 sequence  
and elliptic divisibility sequences.  
%The construction of the elliptic curve associated
%to the Somos 4 sequence is presented as an example.
An interpretation via the theory of integrable systems suggests 
the relation between certain higher order recurrences and 
hyperelliptic
curves of higher genus. Analogous sequences associated with a 
$q$-discrete Painlev\'e I equation are briefly considered.  
\end{abstract}

\section{Introduction}

Question: What's the next number in this sequence: 
$$ 
1,1,2,3,\ldots ?  
$$ 
If you thought the answer is 5, then you guessed wrong! 
Actually the required answer is 7, and it continues  
\beq 
1,1,2,3,7,23,59,314,\ldots,  
\label{somosnum} 
\eeq 
but it's a terrible question, as any four numbers can serve 
as initial data for a recurrence of fourth order or higher. 
Unfortunately this is the sort of question that is routinely 
given to schoolchildren - I noticed it in a recent GCSE exam 
set to 15-year-olds in Britain.  

So if it's not Fibonacci, which comes from a second order 
linear recurrence relation, then how is the sequence 
(\ref{somosnum})  generated? It comes from 
the {\it bilinear} fourth order recurrence 
\beq
\tau_{n+2}=
\frac{
\tau_{n+1}\tau_{n-1}+(\tau_n)^2
}{
\tau_{n-2}
},
\qquad \tau_0=\tau_1=\tau_2=\tau_3=1,
\label{som4}
\eeq
and goes by the name of the Somos 4 sequence   
in Sloane's catalogue \cite{sloane}. 
It turns out that with the special
choice of initial conditions as in (\ref{som4}),
all numerators cancel with their denominators in such a way
that all the terms in the sequence are integers.
 
The remarkable fact is that the sequence (\ref{somosnum}) 
is related to the elliptic curve 
\beq
y^2=4x^3-4x+1.
\label{somcurve}
\eeq
Morever, there is a whole family of sequences defined by 
fourth order bilinear
recurrences, of the form
\beq
\tau_{n+2}\tau_{n-2}=\al \,\tau_{n+1}\tau_{n-1}+\be \,(\tau_n)^2 
\label{bil}
\eeq
with constant coefficients $\alpha$, $\beta$. 
These include a type of integer sequences, known as 
elliptic divisibility sequences,  
\cite{ward1,shipsey}, and their generalizations
\cite{rob,swart}. The connection with elliptic curves is 
a general property of the family of fourth order bilinear
recurrences.  
The following theorem is proved in \cite{melms}.    

\noindent {\bf Theorem.} {\it The general solution
of the quadratic recurrence relation
(\ref{bil}) for $\al\neq 0$ takes the form
\beq
\tau_n=
%(-1)^{n(n-1)/2} [scaled out - signs!]
A\,B^n\frac{\si (z_0 +n\ka )}{
\si (\ka )^{n^2}},
\label{form}
\eeq
where $\ka$ and $z_0$ are non-zero
complex numbers, the constants $A$ and $B$ are given by
\beq
A=\frac{\tau_0}{\si (z_0 )}, \qquad
B=\frac{\si (\ka )\si (z_0 )\,\tau_1}
{\si (z_0+\ka )\,\tau_0},
%\quad\frac{\tau_1}{\tau_0},
\label{consts}
\eeq
and $\si$ denotes the Weierstrass sigma function
associated to the elliptic curve
%$$
$y^2=4x^3-g_2x-g_3$.
%$$
}
%\vspace{.1in}

\noindent The complex constants $z_0,\ka$ should be
considered as points on
the Jacobian of the curve, hence only defined modulo
the period lattice, and together with the invariants $g_2,g_3$
%(the Eisenstein series)
they are determined
uniquely in terms of the constants $\al ,\be$ and the
initial data $\tau_0\neq 0,\tau_1,\tau_2,\tau_3$ specified for the
fourth order recurrence
(\ref{bil}). The precise construction of the curve from
$\al ,\be$ and the
initial data is presented in \cite{melms}. 
Note that in the
case $\al =0$ excluded by the Theorem, the recurrence
(\ref{bil}) decouples into two trivial recurrences for
even and odd terms, so that
$
\tau_{2k}=\tau_0 (\tau_2/\tau_0)^k \be^{k(k-1)/2},
$ and $%\quad
\tau_{2k+1}=\tau_1 (\tau_3/\tau_1)^k \be^{k(k-1)/2}.
$
%}
%\vspace{.1in}

The terms in the Somos 4 sequence (\ref{somosnum}) are given by
formula (\ref{form}) with
$$
\ka =2\omega_1-\int_1^\infty (4t^3-4t+1)^{-\frac{1}{2}}\,dt
=1.859185431 
%=-1.134273216,
%\equiv 2\omega_1+\ka =1.859185430,
$$
$$
z_0=2\omega_3+\int_{-1}^\infty (4t^3-4t+1)^{-\frac{1}{2}}\,dt
=0.204680500+1.225694691i,
%\equiv z_0+2\omega_3
%=0.204680500+1.225694691i ,
%z_0 is complex: hard to get MAPLE to evaluate it.
$$
where the integrals are taken along suitable paths in the
$t$ plane  avoiding the three real roots of the cubic,
and have been evaluated to 9 decimal places using version 8 of the
MAPLE computer algebra package. The real and imaginary half-periods
are $\omega_1=1.496729323$ and $\omega_3=1.225694691i$ respectively.
The other constants appearing in (\ref{form}) are just
$$
\si (\ka )= 1.555836426, \qquad
A= 0.112724016-0.824911686i,
%\frac{1}{\si (z_0)}
$$
$$
B= 0.215971963+0.616028193i.
%\frac{\si (\ka )\si (z_0 )}
%{\si (z_0+\ka )}.
$$
However, in this case the sequence of arguments of
the sigma function can be written
more succinctly as
$$
z_0+n\ka \equiv (2n-3)\hat{z}_0, \qquad
\hat{z}_0=0.929592715+\omega_3,
$$
so that the iterates of the recurrence correspond to the
sequence of points $(2n-3)P$ on the curve (\ref{somcurve}),
where $P=(0,1)$; a referee was helpful in pointing this out. 

There is currently much number theoretical interest in elliptic divisibility
sequences (integer sequences satisfying $\tau_n |\tau_m$ whenever
$n|m$), and in Somos sequences, since apparently they are a
source of large prime numbers \cite{eew, ew2}.

\section{Symplectic structure and integrability}

\setcounter{equation}{0}
By the change of variables 
\beq
f_n=\frac{\tau_{n+1}\tau_{n-1}}{(\tau_n)^2},
\label{fdef}
\eeq
the bilinear recurrence is related to a second order 
difference equation, and so $\tau_n$ is a tau-function. 
 
\noindent {\bf Proposition.} {\it
For $\al\neq 0$ the second order nonlinear map
\beq
f_{n+1}=\frac{1}{f_{n-1}f_n}\left( \al +\frac{\be}{f_n}\right)
\label{nonli}
\eeq
has the general solution
\beq
f_n=\wp (\ka )-\wp (z_0 +n\ka )
\label{desol}
\eeq
with complex parameters $z_0$
and $\ka\neq 0$,
where the $\wp$ is the Weierstrass elliptic function
associated to the curve
\beq
y^2=4x^3-g_2x-g_3,
\label{ecur}
\eeq
and the points $(\wp (\ka ),\wp'(\ka ))$
and $(\wp (z_0 ),\wp'(z_0 ))$ on the curve,
together with the values of the
invariants
$g_2$, $g_3$, are determined
explicitly by the parameters $\al$, $\be$ and the
non-zero initial data $f_0$, $f_1$ of the map
(\ref{nonli}).
}

The second order discrete equation (\ref{nonli})
can be thought of
as a %two-dimensional
mapping of the plane,
$(f_{n-1},f_n)\to (f_n,f_{n+1})$,
with the first integral
$$
\la = \frac{1}{3}\left[
f_n+f_{n+1}+\frac{1}{4\al}\left( f_nf_{n+1}
-\frac{\be}{f_nf_{n+1}}-\frac{\al}{f_n}-
\frac{\al}{f_{n+1}}\right)^2\right].
$$
satisfying $\la (f_n,f_{n+1})=\la (f_{n-1},f_n)$. 
This two-dimensional
mapping is {\it not} symplectic, because
$(f_{n+1})^{-1}\,df_n\wedge df_{n+1}=(f_{n-1})^{-1}\,df_{n-1}\wedge df_n$.
 
To get a symplectic map, we make the transformation to symplectic coordinates 
$(q_n,p_n)$ on the curve (with two-form $dp_n\wedge dq_n$), by setting 
$
\mu=\sqrt{\al}
$
and 
\beq
q_n=\la -f_n, \qquad
p_n=\frac{1}{\mu}f_n^2(f_{n+1} -f_{n-1}).
\label{canon}
\eeq
With these variables the mapping can be rewritten as
\beq
\begin{array}{ccl}
q_{n+1} & = & \Phi_n^2-q_n-\la, \\
p_{n+1} & = & -2\Phi_n^3+(4q_n+2\la )\Phi_n-p_n
, \end{array}
\label{symap}
\eeq
where
$$
\Phi_n=
\frac{p_n-\mu}{2(q_n-\la )}, \, %quad  
\mu (q_n,p_n)=\sqrt{p_n^2+4(\la^3-q_n^3)-g_2(\la -q_n)}. 
$$ 
Note that $\la$ and $g_2$ should be interpreted as constant parameters for 
the 
map (\ref{symap}), while $\mu$ is interpreted as a function 
on phase space. 

The symplectic map (\ref{symap}) is integrable,  with 
first integral 
\beq
g_3(q_n,p_n)=4q_n^3-g_2q_n-p_n^2,
\label{1stint}
\eeq
and has a $2\times 2$ discrete Lax pair \cite{melms}. With a {\it different} 
choice of symplectic structure on the elliptic curve, it is equivalent to 
the one-particle discrete Garnier system in \cite{us1}.

It is natural to wonder how all of the above might
generalize to the case of the
bilinear recurrences
\beq
\tau_{n+g+1 }\tau_{n-g-1} =\sum_{j=0}^{g}\al_{j}\,
\tau_{n+j}\tau_{n-j}, \label{kbil}
\eeq
which for $g\geq 2$ is the natural higher order
analogue of (\ref{bil}), with $g+1$ parameters
$\al_0,\al_1,\ldots,\al_g$. The arithmetic of sequences
defined by such recurrences
is discussed in \cite{recs}.
We are able to state the following 

\noindent {\bf Conjecture. } {\it
The general solution of the
bilinear recurrence (\ref{kbil})
for $g\geq 2$ is given in terms of the Riemann theta-function
associated to an algebraic curve $\Gamma$ of genus
$g$. Using the change of variables (\ref{fdef}),
the bilinear recurrence yields a nonlinear discrete 
equation\footnote{ 
%In the variables $f_j$, 
The recurrence is 
$ 
f_n^{g+1}\prod_{k=1}^{g+1} (f_{n+k}f_{n-k})^{g+1-k}=\al_0+\sum_{j=1}^g\al_j 
f_n^j\prod_{k=1}^j (f_{n+k}f_{n-k})^{j-k}. 
$ 
} 
of order $2g$, which is equivalent to an integrable
symplectic map that linearizes on
$\mathrm{Jac}(\Gamma )$, the Jacobian of
the algebraic curve.
}

\noindent The complete proof of this conjecture is work
in progress with Harry Braden, %\cite{quadr},
but here we briefly outline its main ingredients and explain
why the result is plausible. The result can
be seen as a direct consequence of certain vector addition theorems
for theta-functions obtained by Buchstaber and Krichever \cite{bk}.
However, it turns out that the recurrence (\ref{kbil}) is naturally
related to a hyperelliptic curve $\Gamma$ of the form
\beq
y^2=4x^{2g+1}+\sum_{j=1}^{2g}c_jx^{2g-j},
\label{gspec}
\eeq
(the KdV spectral curves \cite{us1}) 
rather than to an arbitrary curve of genus $g$. Furthermore,
it is most convenient to express the general solution
using the Kleinian sigma function of the curve
(this is equivalent to an expression in theta-functions - see \cite{bel}).
In that case the formula for $\tau_n$ is almost identical
to (\ref{form}) except that $z_0\in\mathrm{Jac}(\Gamma )$
is a vector in the $g$-dimensional Jacobian corresponding to
the reduced divisor $D_0=(P_1-\infty )+(P_2-\infty )+\ldots
+(P_g-\infty )$ of $g$ points on the curve, and $\ka\in\mathrm{Jac}(\Gamma )$
corresponds to the reduced divisor $D=P-\infty $ of another point $P$
(with a derivative of $\sigma$ taken on the theta divisor).

The iterates of the recurrence
correspond to the sequence of divisors $D_0+nD$, which gives
a linear flow $z_0+n\ka$ on $\mathrm{Jac}(\Gamma )$.
The division polynomials for hyperelliptic curves, corresponding
to the special case of multiples of a single point, $nD=n(P-\infty )$,
have been considered by Cantor \cite{cantor}, who gave
examples of these recurrences. Matsutani has also
done some detailed calculations of addition formulae
in genus two
\cite{matsutani}.

\section{An integrable $q$-difference analogue} 

\setcounter{equation}{0}
 
The second order nonlinear difference   
equation (\ref{nonli}) has a non-autonomous, $q$-difference 
version given by 
\beq 
\label{qdiff} 
f_{n+1}f_{n-1}=\frac{\al q^n f_n+\be}{f_n^2}.  
\eeq  
This map satisfies the singularity confinement test 
and has a $4\times 4$ Lax pair \cite{ohta}, as well as a continuum 
limit to the first Painlev\'e equation (PI). Therefore 
(\ref{qdiff}) may be referred to as qdPI. The bilinearization 
of qdPI is achieved by the same formula  
(\ref{fdef}), to obtain the the $q$-difference analogue of 
(\ref{bil}), that is 
\beq
\tau_{n+2}\tau_{n-2}=\al q^n\,\tau_{n+1}\tau_{n-1}+\be \,(\tau_n)^2.  
\label{qbil}
\eeq

It is a remarkable fact that the bilinear equation   
(\ref{qbil}) shares some of the divisibility properties 
of the autonomous recurrences (\ref{bil}), in the sense that 
given constants $\tau_{-2}$, $\tau_{-1}$, $\tau_{0}$, $\tau_{1}$
as initial data, all of the $\tau_n$ for $n\geq 2$ are polynomials 
in $q$ (with coefficients being rational in $\al$, $\be$ and the four initial data).  
So far we have only found an inductive argument for this, but ideally 
$\tau_n$ should be given by a determinant, as for {\it linear} 
$q$-special functions \cite{ismail}. 
When the formula  
(\ref{form}) for the solution to the autonomous recurrence (\ref{bil}
is specialized to rational curves, then there are sequences 
where $\tau_n$ satisfies a linear recurrence. 
However, for (\ref{qbil}) it is possible to prove that 
this can only occur for the special values $q=\pm 1,\pm i$. 

The recurrence relation (\ref{qbil}) is invariant under gauge transformations 
$\tau_n\to \tilde{A}\tilde{B}^n\,\tau_n$, while the transformation 
$\tau_n\to \tilde{C}^{n^2}\tau_n$ means that $\al$ can be rescaled to 1. 
Let us consider the particular sequence for $\al =\be =1$ defined by 
$$ 
\tau_{n+2}=
\frac{
q^n \,\tau_{n+1}\tau_{n-1}+(\tau_n)^2
}{
\tau_{n-2}
},
\qquad \tau_{-2}=\tau_{-1}=\tau_0=\tau_1=1,   
$$ 
which is a non-autonomous version of the Somos 4 sequence (\ref{som4}). 
We would like to make the following observations: 
\begin{itemize} 
\item For $n\geq -2$, $\tau_n=\tau_n(q)$ is  a polynomial in 
$q$ of degree $d_n=n^3/18-n/6  +\frac{2}{9\sqrt{3}}\sin (2\pi n/3)$. 
\item The coefficients of $\tau_n(q)$ are all positive integers; the coefficient of 
$q^{d_n}$ is $2^{[\frac{n+1}{3}]}$. 
\item The roots of $\tau_n(q)$ appear to lie in a circle of maximal 
radius $\sqrt{2}$ in the complex $q$ plane, and coalesce 
to certain special values, 
one of which is $-1.3877957$ to 7 decimal places. 
\end{itemize} 
The rigorous analysis of the properties of the qdPI tau-functions 
will be the subject of future 
work\footnote{During the SIDE 6 meeting, Sasha Bobenko and Vassilios 
Papageorgiou informed me that this should be related to the Laurent 
phenomenon for certain recurrences recently studied by Fomin \& Zelevinsky, 
{\tt arXiv.math.RT/0311493}}. 

\newpage
%\normalsize 
\small 
\noindent {\bf Acknowledgments.}
I am grateful to Graham Everest for introducing me to the
arithmetic of quadratic recurrence sequences, and to
Christine Swart for sending me her thesis \cite{swart}.
Thanks also to Harry Braden and Victor Enolskii for
useful discussions, and to the University of Kent
for supporting
the project
{\it Algebraic curves and functional equations
in mathematical physics} with a Colyer-Fergusson Award.

%\pagestyle{empty} 
%\tiny 

\end{document}